\documentclass[aps,prb,twocolumn,noshowpacs,,superscriptaddress]{revtex4}
\usepackage{graphicx}
\usepackage{graphicx}
\usepackage{dcolumn}
\usepackage{bm}

\begin{document}

\title{Spin-Hall effect: Back to the Beginning on a Higher Level \\
{\small Summary of the APCTP Workshop on the Spin-Hall Effect and Related Issues}\\
{\small Asian Pacific Center for Theoretical Physics, Pohang, South Korea}}
\author{Jairo Sinova}
\affiliation{Department of Physics, Texas A\&M University, College
Station, TX 77843-4242, USA}
\author{Shuichi Murakami}
\affiliation{Department of Applied Physics, University of Tokyo, Hongo, Bunkyo-ku, Tokyo 113-8656, Japan}
\author{Shun-Qing Shen}
\affiliation{Department of Physics,The University of Hong Kong, Hong Kong, China}
\author{Mahn-Soo Choi}
\affiliation{Department of Physics, Korea University, Seoul 136-701, Korea}
\date{\today}
\begin{abstract}
The phenomena of the spin-Hall effect, initially proposed over three decades ago in the context of asymmetric
Mott skew scattering, was revived recently by the proposal of a possible intrinsic spin-Hall effect originating 
from a strongly spin-orbit coupled band structure. 
This new proposal has generated an extensive debate and controversy over the past two years. 
The purpose of this workshop, held at the Asian Pacific Center for Theoretical Physics,
 was to bring together many of the leading groups in this field to resolve such issues and identify
future challenges. We offer this short summary to clarify the now settled issues on some 
of the more controversial aspects of the debate and help refocus the research efforts in new and important avenues.
{\center{\bf Workshop Participants}

I. Adagideli, G. Bauer, M.-S. Choi, Zhong Fang, B. I. Halperin, N. V. Hieu, Jiang-Ping Hu, J. Inoue, H.W. Lee, Minchul Lee, E. Mishchenko, L. Molenkamp, S. Murakami, B. Nikolic, Qian Niu, Junsaku Nitta, M. Onoda, J. Orenstein, C. H. Park, Y.S. Kim, Shun-Qing Shen, D. Sheng, A. Silov, J. Sinova, S. Souma, J. Wunderlich, X. C. Xie, L. P. Zarbo, S.-C. Zhang, Fu-Chun Zhang }
\end{abstract}
\maketitle


\section{Introduction}
The spin Hall effect (SHE) is the generation
in a paramagnetic system of a spin current perpendicular to an applied charge current
leading to a spin accumulation with opposite magnetization at each edge. 
This effect was first predicted over three decades ago by invoking the phenomenology of 
the earlier theories  of the anomalous Hall effect in ferromagnets, which associated its origin to 
asymmetric Mott-skew and side-jump scattering from impurities due to spin-orbit coupling.\cite{Dyakonov:1971_a,Hirsch:1999_a}

Recently the possibility of an intrinsic (dependent only on the electronic structure) SHE has been put forward   \cite{Murakami:2003_a,Sinova:2004_a} predicting the presence of a spin current 
generated perpendicular to an applied electric field in semiconducting systems with strong spin-orbit coupling,
with scattering playing a minor role. This proposal has generated an extensive theoretical debate 
in a very short time motivated by its novel physical concept and potential as a spin injection tool.\cite{LANL} 
The interest has also been dramatically enhanced by recent experiments by two groups
reporting the first observations of the SHE in n-doped semiconductors\cite{Kato:2004_d,Sih:2005_a}  and in 2D hole gases (2DHG).\cite{Wunderlich:2004_a}

These experiments measure directly the spin accumulation induced at the edges of the examples through different optical techniques. On the other hand, most of the early theory has focused on the spin-current generated by an electric field
which would drive such spin-accumulation. In most studies this spin current and its associated conductivity
has been defined as $j_y^z\equiv\{v_y,s_z\}/2=\sigma^{SHE} E_x$. This choice is a natural one but not a unique
one in the presence of spin-orbit coupling 
since there is no continuity equation for spin density as is the case for charge density.
The actual connection between the spin-accumulation and the induced 
spin-current is {\it not} straight forward in the situations where spin-orbit coupling is strong 
and this relation is the focus of current research and one of the key challenges ahead. 

Although two model Hamiltonians with strong spin-orbit coupling have been considered initially,
the p-doped 3D valence band system\cite{Murakami:2003_a} and the 2DEG with Rashba coupling,\cite{Sinova:2004_a} 
the one that has attracted the most attention, perhaps due to its simplicity, is the latter one which has
 the form $H_{\rm R-SO}=\lambda(\sigma_x k_y-\sigma_y k_x)$. 
In such systems, in a clean sample, where the  transport scattering rate
$\tau^{-1}$ is small compared to the spin-orbit  splitting $\lambda  k_F /\hbar$, one finds an intrinsic value  $e/8\pi$for the spin Hall
conductivity, which is valid at finite frequencies  in the range
$\tau^{-1} < \omega < \lambda  k_F / \hbar $, independent of details of
the impurity scattering, in the usual case where both spin-orbit split
bands are occupied. The prediction for the dc spin Hall effect in this
model has been examined and debated extensively.
It was first noticed that contributions to the spin-current from impurity scattering, even 
in the limit of weak disorder, seemed to cancel exactly the intrinsic contribution.\cite{Inoue:2004_a,Mishchenko:2004_a} This lead to speculation
that this cancelation destroys the effect in other model as well. On the other hand, it is now understood through recent
efforts, culminating in this workshop, that such cancelation only occurs for this
\emph{very particular model}, due to the linearity of the spin-orbit coupling and the parabolic
dispersion.\cite{Dimitrova:2004_a,Chalaev:2004_a}

This motivates the title of this summary: After 
our initial excitement and our initial worries that such a beautiful effect may not exist, we are back to the
original proposal but at a higher level of understanding:
that an intrinsic contribution to the SHE in many systems with strong enough 
spin-orbit coupling is present in general.\cite{Murakami:2003_a,Sinova:2004_a} What follows 
is a summary of the issues agreed upon and debated during the open discussion sessions
of the workshop; it is not meant as a summary
of all the topics presented in the workshop. Even though 
feedback from all the speakers in the workshop has been solicited in composing this summary,
any ommisions or unnintentional unbalance is ultimately the responsability of the organizers.
For further information on this workshop and
to view the slides of the talks given and other topics discussed which are not mentioned
here we encourage the reader to visit the workshop website.\cite{site}

\section{Agreement and consensus}
Within the open sessions of this workshop, several key points were discussed and agreement was reached on their 
conclusions. This is an important and intended result of this workshop, to bring together several of the 
leading researchers in the field to clarify the now extensive debate in the literature which can be overwhelming
to a newcomer. 

The agreed upon statements are as follows:
\begin{itemize}
\item[] \emph{The dc spin Hall conductivity, defined through $j_y^z\equiv\{v_y,s_z\}/2=\sigma^{SHE} E_x$, 
does not vanish in general and it includes both intrinsic and non-intrinsic contributions.}

\item[] \emph{The dc spin Hall conductivity for the model Hamiltonian, 
${\cal H_{\rm R}}=\hbar^2 k^2/2m+\lambda (\sigma_x k_y-\sigma_y k_x)$, vanishes in the absence of a magnetic field and spin-dependent scattering, even in the limit of 
weak scattering. This cancellation is due to the particlar relation in this model between the spin dynamics
$d s_y/dt$ and the induced spin-Hall current, i.e. $d s_y/dt=i[{\cal H_R},s_y] \propto j_y^z$, which in a steady
state situation indicates a vanishing spin-Hall current. 
No such relation exists in more complicated models, where the spin-orbit
coupling is not simply linear in the carrier momentum.}
\end{itemize}

The effects of disorder on the induced spin-current, within linear response, 
come in the form of self-energy lifetime corrections and vertex corrections. 
The life time corrections only reduce this induced current 
through a broadening of the bands without affecting its nature. 
On the other hand, vertex corrections have been the source of important debate since 
they make the intrinsic SHE vanish in the Rashba 2DEG system for any arbitrary amount of scattering.\cite{Inoue:2004_a,Mishchenko:2004_a,Chalaev:2004_a} 
For p-type doping in both 3D and 2D hole gases the vertex corrections vanish
in the case of isotropic impurity scattering.\cite{Murakam:2004_a,Bernevig:2004_c,Shytov:2005_a,Khaetskii:2005_a} 
This result is now understood in the context of the specific relation of the spin-dynamics within this 
particular model as stated above.\cite{Dimitrova:2004_a,Chalaev:2004_a} 
This spin-dynamics are linked to the magneto-electric effect producing a homogeneous in-plane spin polarization by an electric field in a Rashba 2DEG.\cite{Edelstein:1990_a,Inoue:2003_a} These results have recently been found to be consistent with numerical treatments of the disorder through exact diagonalization finite size scaling calculations.\cite{Nomura:2005_b,Nomura:2005_a,Sheng:2005_a} 

It is important to point out however that in the mesoscopic regime, where spin Hall conductance 
of finite size systems rather than conductivity of infinite size systems is considered and the finite width can lead to spin-Hall edge states,\cite{Adagideli:2005_a}
the SHE seems  to also be  present and robust against disorder even in the 2DEG Rashba system although 
its link to the bulk regime is still unclear.
\cite{Hankiewicz:2004_b,Nikolic:2004_a,Sheng:2004_a,Adagideli:2005_a}
\section{Semantics}
Given the extensive literature it was deemed useful to agree upon several semantics and notations in order not to 
create confusion from a lack of communication. With this in mind it was agreed that:
\begin{itemize}
\item[]The spin Hall effect is the antisymmetric spin accumulation in a finite width system driven by an
applied electric field.
\item[]The word \emph{intrinsic} is reserved for the intrinsic contribution to the spin-current generated
in the absence of scattering. This contribution can be calculated through the single bubble diagram within
the diagrammatic technique and corresponds to the ac-limit of $\omega \tau
\rightarrow \infty$ where scattering does
not play a role. For example, the intrinsic spin Hall conductivity of 
the Rashba model is $e/(8\pi)$ and for the p-doped valence system it is
$(e/6\pi^2)(k_F^{h.h}-k_F^{l.h.})(1+\gamma_1/(2\gamma_2))$.
\end{itemize}

\section{Future challenges}
\subsection{Theoretical}
Although there is wide agreement within the theoretical community that a
spin Hall effect similar in magnitude to the predicted intrinsic
contribution should occur in p-doped and in mesoscopic samples,
there are still many remaining challenges in order to fully understand this novel 
effect and related effects in spintronics within strongly spin-orbit coupled systems. At the top of the agenda 
seems to be a need to better understand the spin-accumulation induced by the spin-Hall effect at a more quantitative
level and its relation to the spin-current generated. Some of the issues raised during these open session were:
\begin{itemize}
\item[] What is the effect of the scattering on the induced spin-currents and spin coherence in a strongly spin-orbit
coupled system in general and in specific model at a quantitative level (including the sign of the effect
in the several experimental set-ups)? 
\item[] Can the spin-current density seemingly arising from the Fermi sea lead to spin-accumulation and/or spin transport? 
\item[] A clearer understanding of the different contributions and their scaling with respect to disorder (strength,
types, range, etc.) to the induced spin current is needed.
\item[]  How does spin relax in relation to scattering and to the fact that spin is not a conserved quantity in the strongly spin-orbit coupled regime? How does spin relax near the baoundry?
\item[] Is the effect more readily observable at  mesoscopic scales and is there a relation between the mesoscopic and bulk regime? 
\item[] Are there other spin-current definitions which give a clearer picture and can be more readily connected to spin-accumulation?
\item[] There is a need for a full theory of spin-accumulation (and detection) in strongly spin-orbit coupled systems.
\end{itemize}

These are some of the key issues and questions raised but not by all means the only ones that are being considered
in current research. It is important to realize that 
besides the SHE, there is a plethora of effects, linked to spin-transport dynamics in semiconductors, which are important to understand in the context of strongly spin-orbit coupled systems. One in particular is the spin Coulomb drag,\cite{Damico:2002_a} which is an intrinsic friction mechanism between opposite spin populations studied in non-spin-orbit coupled systems, and is important in degenerate systems where electron-electron interactions are relevant. 

\subsection{Experimental}
On of the clear achievements on the spintronics in recent years has been the experimental observation of this
novel effect through optical means.
Spin transport in spin-orbit coupled systems is governed by  characteristic length scales (mean free path, $l=v_F \tau$, spin precession length $l_{so}=\hbar v_F/\Delta_{so}$), time scales (lifetime,$\tau$ , spin coherence time, $\tau_s$) and by the relative strength of spin-orbit coupling, $\Delta_{so}$ and disorder. From these scales it is generally believed that the SHE observed by  Awschalom et al. \cite{Kato:2004_d}
is in the extrinsic regime and the one observed by Wunderlich et al. \cite{Wunderlich:2004_a} in 2DHG is in the intrinsic regime.

Some of the experimental issues raised during the open dicussion session were: 
\begin{itemize}
\item[]A key remaining experimental challenge is the detection of the effect through electrical means which
could lead to actual useful devices. This detection has to be done in coordination with careful realistic
theoretical modeling of particular devices. 
\item[] It is important to understand and model in further detail the effects of edge electric field induced spin-polarization vs. the spin-Hall effect, and the angle dependence of the
luminescence induced in the present set-ups and their relation to the spin magnetization.
\item[] Is it possible to measure spin current in the bulk; i.e. not indirectly through spin accumulation?
\end{itemize}

\section{Outlook}
The past two years have seen a tremendous amount of research achievements and advances in the area of spintronics which 
continuous to generate many novel ideas and phenomena. 
Besides a good and healthy
competitiveness in the field, it has been a field, as it is demonstrated by organizing this conference, which
moves forward in unison to clarify debates rather than allow them to linger for many years,
helping it to move forward to explore interesting new physics. 

As illustrated by the topics debated throughout the workshop, there are many remaining challenges and a 
very healthy outlook of the field, and not just simply of the spin-Hall effect which is a very small part of
the whole  of the spintronics field.
\acknowledgments
The organizers are grateful for the sponsorship of the Asian Pacific Center for Theoretical Physics and 
the National Science Foundation (OISE-0527227) which have made this workshop possible.

\begin{thebibliography}{27}
\bibitem[{LAN()}]{LANL}
\bibinfo{note}{There are over 180 pre-prints posted in the last two years in
  the LANL archives focused on this effect.}
\end{thebibliography}

\end{document}